# Regulation of oxygen vacancy types on $SnO_2$ (110) surface by external strain

Z. H. Zhou, Y.M. Min, X. X. Liu, J. Q. Ding, L. Z. Liu[a]

*Collaborative Innovation Center of Advanced Microstructures; National Laboratory of Solid State Microstructures, Nanjing University, Nanjing 210093, PR China*

## *Abstract*

In tin dioxide nanostructures, oxygen vacancies (OVs) play an important role in their optical properties and thus regulation of both OV concentration and type via external strain is crucial to exploration of more applications. First-principle calculations of $SnO_2$ (110) surface disclose that asymmetric deformations induced by external strain not only lead to its intrinsic surface elastic changes, but also result in different OV formation energy. In the absence of external strain, the energetically favorable oxygen vacancies (EFOV) appear in the bridging site of second layer. When -3.5% external strain is applied along y direction, the EFOV moves into plane site. This can be ascribed that the compressed deformation gives rise to redistribution of electronic wave function near OVs, therefore, formation of newly bond structures. Our results suggest that different type OVs in $SnO_2$ surface can be controlled by strain engineering.

_______________________________

[a] Corresponding authors.   Email: lzliu@nju.edu.cn (L. Z. Liu)

## 1. Introduction

Tin oxide (SnO$_2$) is one of the attractive functional materials because of potential applications of biophysics, gas sensing, catalysis, and batteries.[1-5] However, their properties depend heavily on the nanostructure morphology as well as electronic structure, which is modified by intrinsic oxygen vacancies (OVs).[6,7] Development of optoelectronic encompassing bulk SnO$_2$ materials, for instance, has been hampered by its dipole forbidden nature.[8,9] Fortunately, owing to OVs existence, changes in symmetry of nanostructures related with surface states may allow direct gap transitions rendering quantum-confined photoluminescence possible. Therefore, different nanostructures such as nanowires, nanoribbons, tetragonal microcubes and nanocrystals have been fabricated,[10-14] and their electronic, phonon and optical characteristics, and potential applications have been explored recently, some fundamental issues pertaining to the novel optical properties of SnO$_2$ nanostructures are still unclear due to the inherent structural complexity. A detailed analysis to previous researches discloses that OVs concentration and types play a decisive role in modification of physical performance. Our previous Raman spectral analysis reveals that the Raman mode behavior in SnO$_2$ nanostructures is closely related to the OV types.[15] The results aspire us to investigate the relation between electronic structure and OV distribution. Three oxygen-vacancy-related photoluminescence peaks at 430, 501 and 618 nm are observed, and Raman scattering and density functional calculation indicate that they originate from in-plane, sub-bridging and bridging OVs, respectively.[16] Those results further confirm that OVs types are crucial to their intrinsic attributes, thus a systematic investigation of OVs formation on nanostructure surface not only elucidates the origin of newly physical behavior but also imparts information about the surface structure necessary for the development of future catalytic and sensing applications. However, the complexity of nanostructures and lack of systematic theoretical assessment make it challenging. External strain can be easily implemented by introducing a particular substrate in the fabrication of the $SnO_2$ nanostructures and the thin films. The responses of nanostructures to the external strain are determined by its mechanical properties, which are strongly influenced by the OV structures. So it becomes an effective way to regulate the electronic structure and superficial elasticity and if the effect of external strain on formation energy of different type OVs can be adjusted, OVs can be in fact be manipulated by strain engineering. Finally, physical performance of OV relevance in $SnO_2$ nanostructures surface can be modulated by this technology.   However, the understanding of strain-dependent OV formation mechanism in $SnO_2$ surface is quite limited so far.

In this work, the $SnO_2$ (110) surface with different type OVs is chosen as the model system to study the relationship between external strain and energetically favorable OVs.  The theoretical derivation shows that asymmetric deformation induced by uniaxial strain not only regulates the surface stress and surface elastic constants but also tunes the type of energetically favorable oxygen vacancy(EFOV), thereby suggesting that strain engineering is an ideal way to control OVs formation.

## 2. Theoretical methods

The theoretical assessment is based on the density functional theory in Perdew-Burke-Ernzerhof (PBE) generalized approximation (GGA), using the CASTEP package code with projector augmented wave pseudopotenials.[17-19] The plane-wave energy cutoff of 750 eV is used to expand the Kohn-Sham wave functions and relaxation is carried out until convergence tolerances of 1.0 x 10$^{-5}$ eV for energy and 0.001Å for maximum displacement are reached.   The vacuum space is at least 13Å, which is large enough to avoid the interaction between periodical images. The Monkhorst-Pack *k*-point meshes (in two-dimensional Brillion zone) of 10x4x1 and 3x1x1 and 1x3x1 for (1x1) and (1x4) and (4x1) slab of SnO$_2$(110) surface, which has been tested to converge.   A typical schematic of the side and top view of rutile SnO$_2$ surface is displayed in Fig. 1(a) and 1(b) in which 5 different kinds of Oxygen atom are characterized by In-plane I ,In-Plane II, Bridging I ,Bridging II and Bridging III.

Surface energy is a critical parameter for SnO$_2$ surface property, which directionally determine its structural stability. The formula for calculating surface energy can be written as:

$$E_S = \frac{E_1 - E_2}{2A} \tag{1}$$

where E$_1$ is the energy of the crystal with two surface, E$_2$ is the energy of the bulk with the same amount of atoms, and A is the area of the surface. The calculated surface energy of (1x1) slab is 1.34 J/m$^2$, which is quite agreement with results of other groups[20-22].

According to the surface symmetry, the different types vacancies in different depth can investigated and labeled as Bridging I, Bridging II, Bridging III, In-Plane I, and

In-Plane II, respectively, as shown in Fig. 1(b). The vacancy formation energy (EOV) can be expressed as:

$$EOV = E_{vac} - E_{st} + 1/2 E_{O2} \tag{2}$$

where $E_{O2}$ is the energy of an isolated $O_2$ molecule (which is -855.32 eV in our calculation). $E_{st}$ and $E_{vac}$ stand for the energy of the stoichiometric structure and that of the same structure except containing an oxygen vacancy, respectively. Then we define the applied strain rate in x and y direction: $\alpha_x$ and $\alpha_y$

$$\alpha_x = \frac{x - x_0}{x_0} \tag{3}$$

$$\alpha_y = \frac{y - y_0}{y_0} \tag{4}$$

Where x and y are the lengths of the supercell in the x and y direction after deformation, and $x_0$ = 2.99 Å and $y_0$ = 6.48 Å are the lengths of the supercell in the x and y direction before deformation.

Next, we analyze the influence of OVs on the surface elasticity of $SnO_2$(110) surface. The surface stress tensor and surface elasticity constants can be defined as[23]:

$$\sigma_{\alpha\beta} = \frac{\partial \gamma_L}{\partial \epsilon_{\alpha\beta}} \tag{5}$$

$$S_{\alpha\beta\alpha'\beta'} = \frac{\partial \sigma_{\alpha\beta}}{\partial \epsilon_{\alpha'\beta'}} \tag{6}$$

here $\gamma_L$ is the Lagrangian surface energy per unit area in an unstrained state and $\epsilon_{\alpha\beta}$ is the strain tensor. The variation of surface elastic constants can be written as:[23]

$$\Delta S_{\alpha\beta\alpha'\beta'} = S'_{\alpha\beta\alpha'\beta'} - S_{\alpha\beta\alpha'\beta'} \tag{7}$$

Where $S'_{\alpha\beta\alpha'\beta'}$ is the surface elastic constants after introducing the vacancy, $S_{\alpha\beta\alpha'\beta'}$ is the elastic constants of perfect surface. The variation of surface elastic constants of SnO$_2$ (110) surface with different type OVs as functions of $\alpha_x$ and $\alpha_y$ is discussed in detailed.

## 3. Results and discussion

The lattice constant of prefect SnO$_2$ (110) surface is $x_0$= 2.99Å and $y_0$= 6.48 Å, then geometry optimization for a primitive (1x1) slab is carried out by removing different site oxygen atoms. The Fig. 1(a) and 1(b) present a side and a top view of the SnO$_2$ (110) surface with different type OVs. In order to break lattice symmetry, the formation energy of different OV types is changed by external strain in distinctive directions.

The oxygen vacancy formation energy (EOV) of (1x1) SnO$_2$ surface as functions of external strain is calculated and shown in Fig. 2. When the external strain is applied along x direction [as shown in Fig. 2(a)], we can see that the EOV increase slowly with compressed deformation from $\alpha_x$ = 5% and reached an maximum value at $\alpha_x \approx -2.3\%$, then decreased sharply. This behavior is different from the issues of Fig. 2(b), as shown that the deformation along y direction only leads to an linear increase of EOV. Those obvious behavioral difference induced by external strain is strongly relevant to their asymmetric lattice structure, because that smaller lattice constant along x direction make electronic wave function overlap and develop into more stable statues. Without external strain, the Bridging I OV and In-Plane IOV is more energetically than Bridging II OV, which is consist with experimental conclusion. This can be easily understood that Bridging I OV and In-Plane IOV exposed to the vacuum directly are more easily divorced from surface. while the internal oxygen atom, such as bridging II site, require more energy to get rid of for the coulomb potential of other atoms. In Fig . 2(a), the EOV of Bridging I OV at $\alpha_x$ = -5 % become 5.39 eV and smaller than that of In-Plane I OV (5.43 eV).   This phenomenon also happen into the structure induced by y direction deformation ($\alpha_y$> 2.5 %) [see Fig .2 (b)].   Those calculations suggest that OV formation types can be changed in deformed surface, though the energy difference is slight.

In application, the OV generally appear at several top layers of $SnO_2$ nanostructure surface, thus the depth different type OVs as functions of external strain along x and y direction are also calculated and shown in Fig. 2(c) and 2(d). To begin with, the values of EOV in depth region are larger than that of shallow region, this can be ascribed to surface potential feature.   Moreover, the changes of EOV with external strain along x and y direction are significantly different, the deformation along y direction cause remarkable EOV changes.   The curve of In-Plane II OV can cross with Bridging III line at $\alpha_y$= -0.5 % and then intersects into Bridging II OV line at $\alpha_y$ = -3.4 %.   That means that the formation order of different type OV can be regulated by external strain along y direction. In addition, the OV types are strongly related with asymmetric deformation and distributed depth, and changes in electronic structure are responsible for this complicated behavior.

To describe physical process induced by deformation, the surface elasticity constants of $SnO_2$ surface with In-Plane IIOV, Bridging II OV, and Bridging III are calculated and shown in the Table. I. and Table. II. From the data, we can find that the In-plane II type OV can the most dramatically affect the surface elasticity constants, and the Bridging III OVs has the least influence on its surface elasticity constants.  What's more, the relation between variation of surface elastic constants and strain rate α is not linear. Those results further indicate that the EOV are not changed lineally, because it is determined by surface elasticity constants.

To visibly display the electron redistribution, the electron density of SnO$_2$ (1x1) surface with In-plane II OV at $\alpha_y$= -3.5 % and $\alpha_y$ = 3.5 % external strain are shown in Fig. 3(c) and 3(d), respectively. As we seen in Fig. 3(c), the compressed deformation make the electron cloud overlaps sufficiently in the In-Plane II OV region and form an weak Sn-Sn bond via $d$ electron hybridization. Detailed population analyses indicate that the interaction between strain and In-Plane OV will exert impact on the atom and band around the vacancy. The length of the Sn-1 –Sn-2 bond decline from 3.741 Å to 3.349Åwhile strain rate $\alpha_y$ changes from 3.5 % to -3.5 % , simultaneously Mulliken Charge of the Sn-1and Sn-2 shift from 2.02e and 1.27e respectively to 1.81e and 1.47 e respectively . Furthermore, the electron cloud begin to overlap for the formation of the more stable Sn-1-Sn-2 bond when the crystal is compressed [see Fig.3.(c)].

Considering the symmetry of surface structure, the supercell of (4x1)  [see fig.3 (a)]and (1x4) slab [see Fig.3(b)]with different OVs in the first layer are also calculated and shown in the Fig. 4(a) and 4(b), respectively. For the (1x4) slab, the formation energy of In-Plane I OV also can be reduced with compressed deformation, and become smaller than that of Bridging I OV while$\alpha_y$< -2.5 %. But this transition cannot occur in the system with Bridging II OV thereby suggesting that strain is useless to control Bridging II OV formation. For the (4x1) slab, although the EOV of In-Plane I and Bridging I OV remain no across down to $\alpha_y$ = -5 % [see Fig. 4(a)], it is possible to change the EOV of those two OVs through compressed deformation. Fig. 4(a) indicate that for larger external compression ($\alpha_y$< -5 %)  the In-Plane I OV may be more energetically favorable than Bridging I OV . Those results further imply that the strain engineering play an important role in OV formation.

## 4. Conclusions

In summary, in the $SnO_2$ surface structure, different type OV formation can be controlled by asymmetric deformation. When -3.5% external strain is applied along y direction, the EFOV moves into plane site from bridging site[see Fig 2.(d)]. Theoretical calculations disclose that asymmetric deformations induced by external strain can lead to electron density redistribution and intrinsic surface elastic change, therefore causes the change of formation energy of different type OVs, finally regulates the OV types. Since the external strain can be easily applied by means of a specific substrate in fabrication, the strategy to apply external strain in engineering of $SnO_2$ OV types is promising .

## Acknowledgements

This work was supported by National Natural Science Foundation (Nos. 11404162), partial support was also from PAPD and Natural Science Foundations of Jiangsu Province(No.BK20130549). We also acknowledge computational resources of High Performance Computing Center of Nanjing University.

# References

[1] S. Brovelli, A. Chiodini, A. Lauria, F. Meinardi, A. Paleari, Phys. Rev. B **73**, 073406 (2006).

[2] H. T. Chen, S. J. Xiong, X. L. Wu, J. Zhu, J. C. Shen, Paul K. Chu, Nano. Lett. **9**, 1926 (2009).

[3] Y. J. Chen, L. Nie, X. Y. Xue, Y. G. Wang, T. H. Wang, Appl. Phys. Lett. **88**, 083105 (2006).

[4] R. Asahi, T. Morijawa, T. Ohwaki, K. Aoki, Y. Taga, Science **293**, 269 (2001).

[5] S. O. Kucheyev, T. F. Baumann, P. A. Sterne, Y. M. Wang, T. Buuren, A. V. Hamza, L. J. Terminello, T. M. Willey, Phys. Rev. B **72**, 035404 (2005).

[6] C. Killic, A. Zunger, Phys. Rev. Lett. **88**, 095501 (2002).

[7] L.Z. Liu, X.L. Wu, J.Q. Xu, T.H. Li, J. C. Shen, Paul K. Chu, Appl. Phys. Lett. **100**, 121903 (2012).

[8] F. Arlinghaus, J. Phys. Chem. Solids **35**, 931 (1974).

[9] G. Blattner, C. Klingshrin, R. Helbig, Solid Stat Commun. **33**, 341 (1980).

[10] O. Lupan, L. Chow, G. Chai, H. Heinrich, S. Park, A. Schulte, J. Cryst. Growth **311**, 152 (2008).

[11] J. Q. Hu, Y. Bando, Q. L. Liu, D. Golberg, Adv. Funct. Mater. **13**, 439 (2003).

[12] E. J. H. Lee, C. Ribeiro, T. R. Giraldi, E. Longo, E. R. Leite, J. A. Varela, App. Phys. Lett. **13**, 1745 (2004).

[12] E. J. H. Lee, C. Ribeiro, T. R. Giraldi, E. Longo, E. R. Leite, J. A. Varela, App. Phys. Lett. **13**, 1745 (2004).

[13] S. H. Luo, P. K. Chu, W. L. Liu, M. Zhang, C. L. Lin, App. Phys. Lett. **88**, 183112 (2006).

[14] L. Z. Liu, X. L. Wu, T. H. Li, S. J. Xiong, H. T. Chen, P. K. Chu, App. Phys. Lett. **99**, 251902 (2011).

[15] L. Z. Liu, X. L. Wu, T. H. Li, J. C. Shen, P. K. Chu, J. Raman. Spectra. **43**, 1423 (2012).

[16] L. Z. Liu J. Q. Xu, X. L. Wu, T. H. Li, J. C. Shen. P. K. Chu, App. Phys. Lett. **102**, 031916 (2013).

[17] B. G. Pfrommer, M. Cote, S. G. Louie, M. L. Cohen, J. Comput. Phys. **131**, 233 (1997).

[18] M. A. Maki-Jaskari and T. T. Rantala, Phys. Rev. B **64** 075407 (2001).

[19] G. Cicero, A. Catellani, G. Galli, Phys. Rev. Lett. **93**, 016102 (2004).

[20] P. A. Mulheran and J. H. Harding, Modell. Simul. Mater. Sci. Eng. **1**, 39(1992).

[21] B. Slater, C. Richard, A. Catlow, D. H. Gay, D. E. Williams, V. Dusastre, J. Phys. Chem. B. **103**, 10644 (1999).

[22] W. Bergermayer and I. Tanaka, App. Phys. Lett. **84**, 909 (2004).

[23] D. J. Shu, S. T. Ge, M. Wang, N. B. Ming, Phys. Rev. Lett. **106**, 039902 (2011).

**Table captions:**

Table. I : Variation of the elastic constants $\Delta S_{1111}$ under different strain rate in x direction. Table. II.: Variation of elastic constants $\Delta S_{2222}$ under different strain rate in y direction.

| $\Delta S_{\alpha\beta\alpha'\beta'}$ (GPa) | $\Delta S_{1111}$ $\alpha_x = -2.5\%$ | $\Delta S_{1111}$ $\alpha_x = 0\%$ | $\Delta S_{1111}$ $\alpha_x = 2.5\%$ |
|---|---|---|---|
| Bridging II | -31.073 | -41.733 | -22.050 |
| In-Plane II | -34.158 | -41.594 | -35.101 |
| Bridging III | -25.578 | -18.474 | -26.737 |

Table. I.

| $\Delta S_{\alpha\beta\alpha'\beta'}$ (GPa) | $\Delta S_{2222}$ $\alpha_y = -2.5\%$ | $\Delta S_{2222}$ $\alpha_y = 0\%$ | $\Delta S_{2222}$ $\alpha_y = 2.5\%$ |
|---|---|---|---|
| Bridging II | 9.384 | -39.318 | -29.720 |
| In-Plane II | -48.086 | -66.262 | -65.701 |
| Bridging III | 6.625 | -6.172 | -6.547 |

Table. II.

## Figure captions

Fig. 1(a) $1 \times 1 SnO2(110)$ models and the axis coordinate .The atom of red color represents the Oxygen atom and the grey one represents Tin atom. (b) The front-view of the model and the 5 kinds of the Oxygen atom .

Fig. 2(a) EVO varies with the strain in x direction in the first layer in $1\times 1$ model. (b) EVO varies with the strain in y direction in the first layer. in $1\times 1$ model. (c) EVO varies with the strain in x direction in the second layer in $1\times 1$ model. (d) EVO varies with the strain in y direction in the second layer in $1\times 1$ model.

Fig.3. (a) $4\times 1$ model. (b) $1\times 4$ model. (c) The electron density isosurface of the crystal under $\alpha_y$=-0.035.    (d) The electron density isosurface of the crystal under $\alpha_y$=0.035. The isovalue is 0.2.

Fig.4.(a) EVO varies with the strain in y direction in the first layer in $4\times 1$ model.(b) EVO varies with the strain in y direction in the first layer in $1\times 4$model.

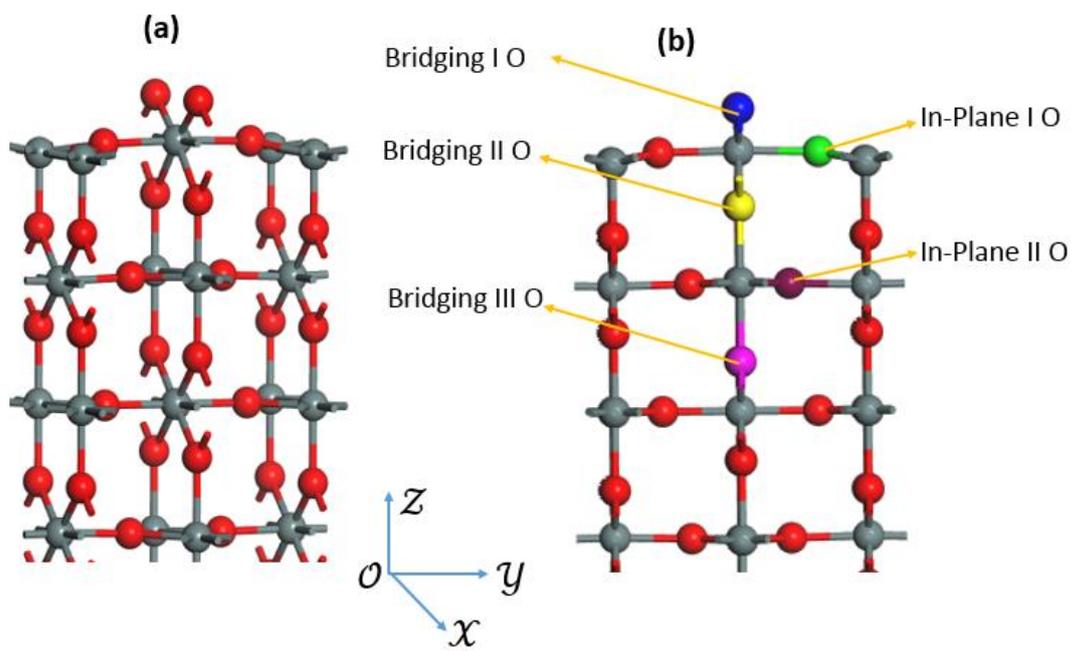

Figure 1

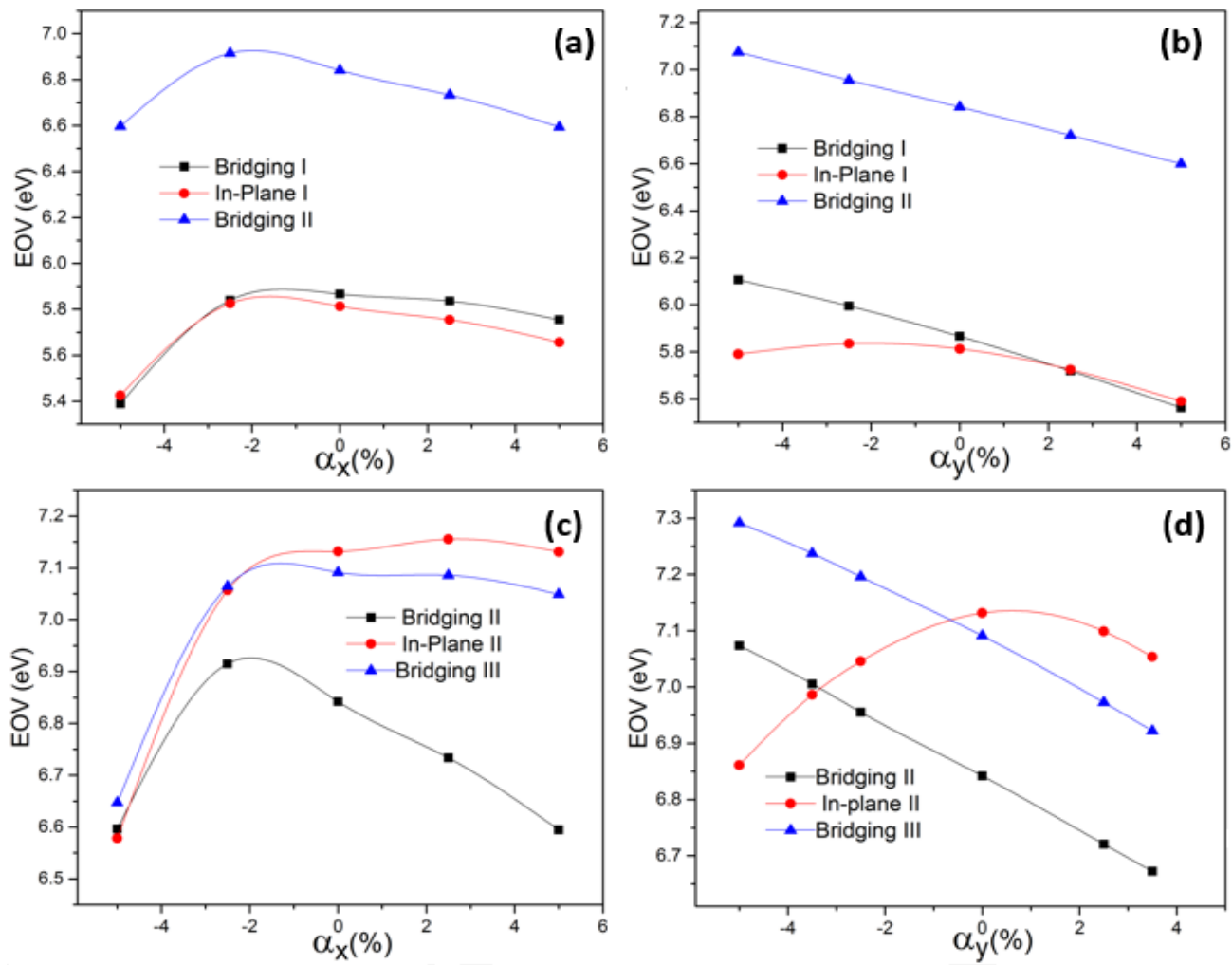

Figure 2

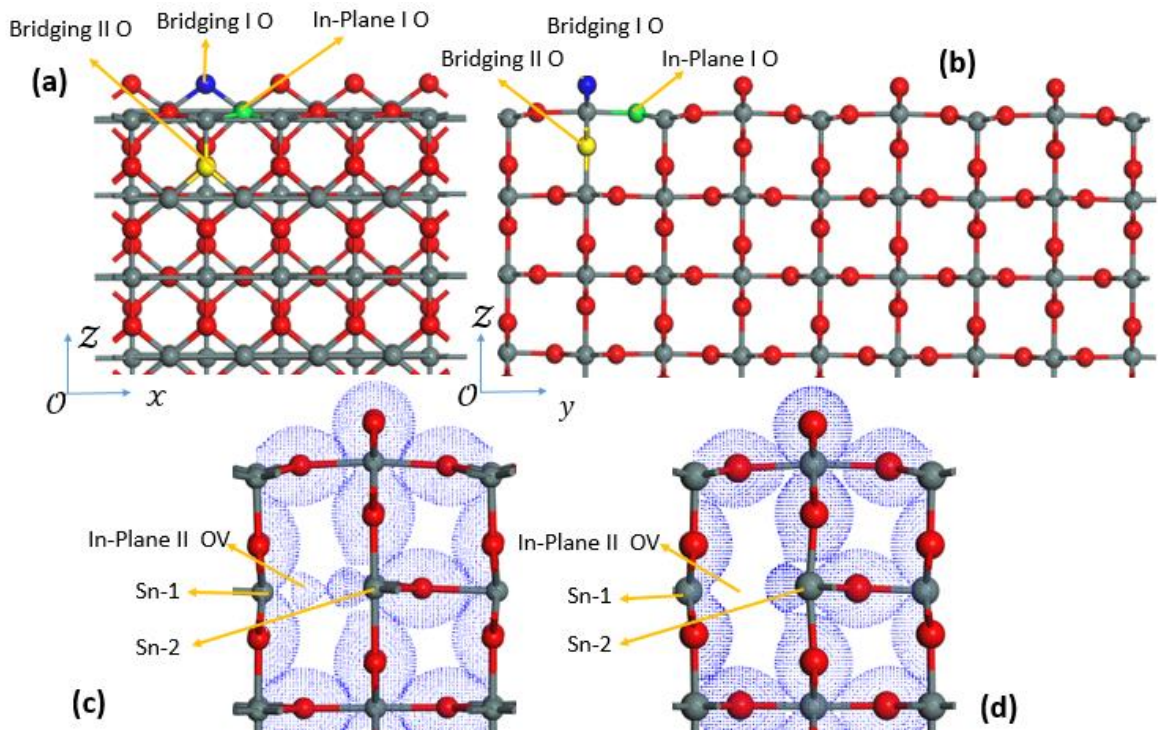

Figure 3

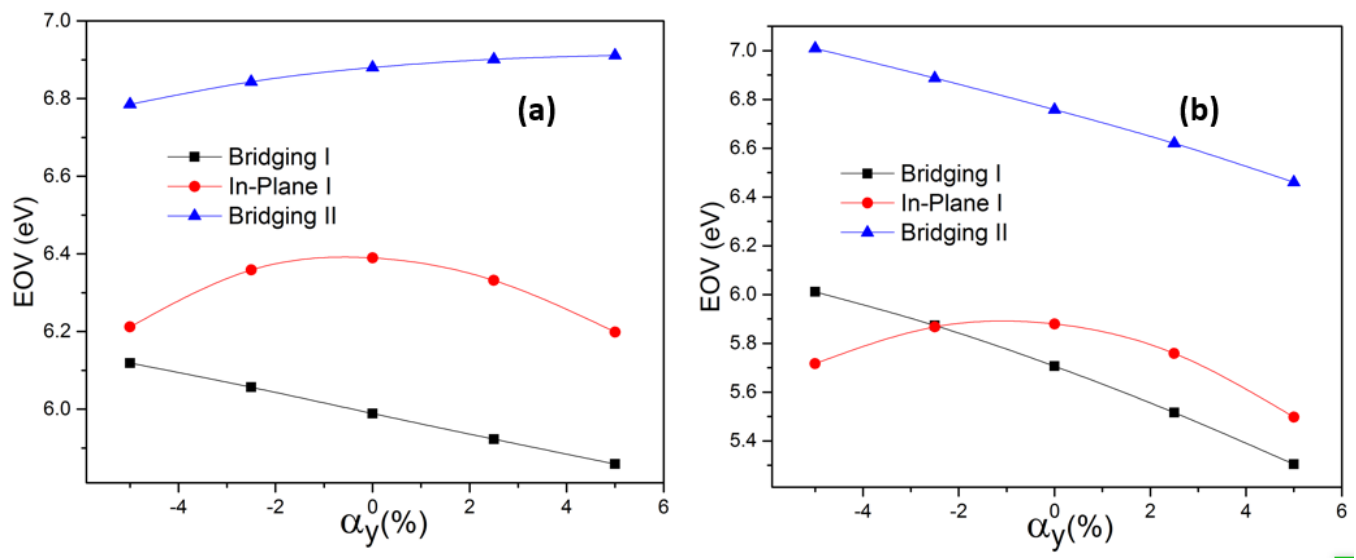

Figure 4